\documentclass[times,twocolumn]{aastex631}

\newcommand{\simless}{\mathbin{\lower 3pt\hbox
     {$\rlap{\raise 5pt\hbox{$\char'074$}}\mathchar"7218$}}}
\newcommand{\simgreat}{\mathbin{\lower 3pt\hbox
     {$\rlap{\raise 5pt\hbox{$\char'076$}}\mathchar"7218$}}}

\newcommand{\ugc}[1]{}

\newcommand{\ahg}{\textcolor{magenta}}

\usepackage{amsmath }
\usepackage{tablefootnote}

\graphicspath{{./}{figures/}}

\accepted{by The Astrophysical Journal}

\shorttitle{Sample article}
\shortauthors{Hales et al.}

\begin{document}

\title{Discovery of an accretion streamer and a slow wide-angle outflow around FU~Orionis}

\author[0000-0001-5073-2849]{A.S. Hales}

\affiliation{National Radio Astronomy Observatory, 520 Edgemont Road, Charlottesville, VA 22903-2475, United States of America}
\affiliation{Joint ALMA Observatory, Avenida Alonso de C\'ordova 3107, Vitacura 7630355, Santiago, Chile}
\affiliation{Millennium Nucleus on Young Exoplanets and their Moons (YEMS), Chile}

\author[0000-0002-9959-1933]{A. Gupta}
\affiliation{European Southern Observatory, Karl-Schwarzschild-Str. 2, 85748 Garching bei München, Germany}

\author[0000-0003-3573-8163]{D. Ru\'iz-Rodr\'iguez}
\affiliation{Joint ALMA Observatory, Avenida Alonso de C\'ordova 3107, Vitacura 7630355, Santiago, Chile}

\author[0000-0001-5058-695X]{J. P. Williams}
\affiliation{Institute for Astronomy, University of Hawaii, Honolulu, HI 96816, USA }

\author[0000-0003-2953-755X]{S. Pérez}
\affiliation{Departamento de Física, Universidad de Santiago de Chile, Av. Victor Jara 3659, Santiago, Chile}
\affiliation{Millennium Nucleus on Young Exoplanets and their Moons (YEMS), Chile}
\affiliation{Center for Interdisciplinary Research in Astrophysics and Space Exploration (CIRAS), Universidad de Santiago de Chile, Chile}

\author[0000-0002-2828-1153]{L. Cieza}
\affiliation{Instituto de Estudios Astrof\'isicos, Facultad de Ingenier\'ia y Ciencias, Universidad Diego Portales, Av. Ejercito 441, Santiago, Chile}
\affiliation{Millennium Nucleus on Young Exoplanets and their Moons (YEMS), Chile}

\author[0000-0003-4907-189X]{C. Gonz\'alez-Ruilova}
\affiliation{Instituto de Estudios Astrof\'isicos, Facultad de Ingenier\'ia y Ciencias, Universidad Diego Portales, Av. Ejercito 441, Santiago, Chile}
\affiliation{Millennium Nucleus on Young Exoplanets and their Moons (YEMS), Chile}

\author[0000-0002-3972-1978]{J. E. Pineda}
\affiliation{Max-Planck-Institut für extraterrestrische Physik, Giessenbachstrasse 1, D-85748 Garching, Germany}

\author[0000-0001-6267-2820]{A. Santamaría-Miranda}
\affiliation{European Southern Observatory, Alonso de C\'ordova 3107, Casilla 19, Vitacura, Santiago, Chile}
\affiliation{Departamento de Astronom\'ia, Universidad de Chile, Camino El Observatorio 1515, Las Condes, Santiago, Chile}

\author[0000-0002-6195-0152]{J. Tobin}
\affiliation{National Radio Astronomy Observatory, 520 Edgemont Road, Charlottesville, VA 22903-2475, United States of America}

\author[0000-0002-3354-6654]{P. Weber}
\affiliation{Departamento de Física, Universidad de Santiago de Chile, Av. Victor Jara 3659, Santiago, Chile}
\affiliation{Millennium Nucleus on Young Exoplanets and their Moons (YEMS), Chile}
\affiliation{Center for Interdisciplinary Research in Astrophysics and Space Exploration (CIRAS), Universidad de Santiago de Chile, Chile}

\author[0000-0003-3616-6822]{Z. Zhu}
\affiliation{Department of Physics and Astronomy, University of Nevada, Las Vegas, 4505 South Maryland Parkway, Las Vegas, NV 89154-4002, USA}
\affiliation{Nevada Center for Astrophysics (NCfA), University of Nevada, Las Vegas, NV, USA}

\author[0000-0002-5903-8316]{A. Zurlo}
\affiliation{Instituto de Estudios Astrof\'isicos, Facultad de Ingenier\'ia y Ciencias, Universidad Diego Portales, Av. Ejercito 441, Santiago, Chile}
\affiliation{Millennium Nucleus on Young Exoplanets and their Moons (YEMS), Chile}

\begin{abstract}

We present ALMA 12-m, 7-m \& Total Power (TP) Array observations of
the FU~Orionis outbursting system, covering spatial scales ranging
from 160 to 25,000~au. The high-resolution interferometric 
data reveals an elongated $^{12}$CO(2--1) feature previously
observed at lower resolution in $^{12}$CO(3--2). Kinematic modeling
indicates that this feature can be interpreted as an accretion
streamer feeding the binary system. The mass infall rate provided by
the streamer is significantly lower than the typical stellar
accretion rates (even in quiescent states), suggesting that this
streamer alone is not massive enough to sustain the enhanced
accretion rates characteristic of the outbursting class
prototype. The observed streamer may not be directly linked to the
current outburst but rather a remnant of a previous, more massive
streamer that may have contributed enough to the disk mass to render
it unstable and trigger the FU~Ori outburst. The new data detects,
for the first time, a vast, slow-moving carbon monoxide molecular
outflow emerging from this object.  To accurately assess the outflow
properties (mass, momentum, kinetic energy), we employed
$^{13}$CO(2--1) data to correct for optical depth effects. The
analysis indicates that the outflow corresponds to swept-up material
not associated with the current outburst, similar to slow-molecular
outflows observed around other FUor and Class~I protostellar
objects.

\end{abstract}


\section{Introduction}\label{intro}

The flaring of FU~Orionis in 1936 \citep{Hoffleit1939} marked the
discovery of a new phase in low-mass star formation. FU~Ori increased
its brightness by over 5 magnitudes within a year
\citep{wachmann1954}. \citet{herbig1966} noted that the brightening
could not be explained in terms of a nova event or by sudden changes
in extinction, and that it was most likely due to {\it`some
    phenomenon of early stellar evolution.'}  \citet{hartmann1985}
argued that the increased luminosity could be explained by  a
rapid onset of accretion from a rotating accretion disk (i.e. an
accretion outburst). In the accretion disk model, the energy released
by the accreting gas heats up the disk up to several thousand K,
reproducing the increase in luminosity, observed spectral energy
distribution (SED), and peculiar spectral properties such as the
apparent wavelength-dependent stellar spectral type which can be
explained by absorption from different regions of superheated disk
atmosphere \citep{hartmann1996,hartmann2016}.

For decades FU~Ori was the only known star of its class, but
the progressive discovery of more outbursting sources suggested that
these are not isolated cases, but a common, yet short-lived, phase in
star formation. Newly discovered outbursting stars showed diverse
outburst properties forcing the definition of sub-classes: FUors
(named after FU~Ori) have high-amplitude ($>$ 5 mag), long-lived
(years to decades) optical bursts \citep{herbig1966}, whereas EXors
(misnamed after EX~Lupi) have low amplitude (2--4 mag) and shorter
(days to months) periods of enhanced accretion \citep[e.g.][for a
  review]{audard2014}. The advent of systematic photometric surveys
 increased the discovery rate of new eruptive objects
\citep[e.g.][]{kun2014}.  Milky Way surveys indicate that 3-6~$\%$ of
Class I protostars show eruptive behavior in a 4-year span and that
eruptive events are more common in Class I objects than in Class II
\citep{contreras2017}.

As new sources are discovered, the heterogeneous nature of their
characteristics presents a significant challenge in terms of
classification and understanding. \citet{Fischer2022} discusses the
difficulties encountered in distinguishing between various types of
outbursts, such as FU~Orionis-like events and EX Lupi-like
events. These challenges arise due to the overlapping characteristics
exhibited by these phenomena and the limitations imposed by the
available observational data. Recent advancements in time-dependent,
multiwavelength, and spectroscopic analyses suggest a continuum of
diverse phenomena rather than distinct and separable classes. This
emphasizes the need for comprehensive and multifaceted approaches to
understand the complexity of these outburst events fully.

Accretion outbursts are now believed to be central to the formation of
low-mass stars. The well-known Luminosity Problem \citep{kenyon1990}
arises from the fact that the measured luminosities of low-mass
protostars are smaller than expected by the quiescent
(non-outbursting) mean accretion rates (10$^{-6}$
M$_{\odot}\,$yr$^{-1}$). The problem is mitigated if the accretion
rates are variable, with stars accreting an important fraction of their mass
during several bursts of enhanced accretion during the Class~I phase
\citep{evans2009,dunham2010} while spending most of their time in
lower quiescent accretion rates than  predicted (10$^{-8}$
M$_{\odot}\,$yr$^{-1}$). Submillimeter monitoring of star-forming
regions indicates that episodic accretion may occur even earlier, at
the Class~0 stage \citep{safron2015,johnstone2018}.

The physical mechanisms responsible for the outburst remain a matter
of debate. A funneling mechanism in which matter from an envelope is
accumulated in the accretion disk and released episodically could
explain the outbursts \citep{zhu2009}. Thermal instabilities can also be
responsible for the piling up and sudden release of inner disk
material \citep{bell1994}. Additionally, the formation and fragmentation of spiral waves, driven by gravitational instabilities, have been suggested to trigger these accretion outbursts \citep[see][for a review]{2016ARA&A..54..271K}. Observations of spiral structures due to gravitational instabilities, such as those seen in the massive disk of Elias 2-27 \citep{2016Sci...353.1519P} and the recently observed fragmenting spirals around the V960 Mon system after its FUor-like outburst in 2014 \citep{weber2023}, support this theory.

Despite their significance in our understanding of low-mass star formation, the exact causes of these outbursts remain elusive. The dense star-forming environments of these young circumstellar disks offer another possible explanation. It is proposed that the sudden outbursts could be linked to interactions within binary systems or encounters with passing stars, which are quite common in these crowded regions \citep{bonnell1992,reipurth2004}. Additionally, the multiplicity of stars and the substantial cross-sectional area of these disks can foster interactions with passing stars \citep[star-disk flybys;][] {pfalzner2008,cuello2019} as well as disk-disk interactions which can excite the disks turning them unstable \citep{munoz2015}.

Various FUor-like protostars have been identified as binary
companions, including notable cases such as L1551 IRS5, RNO 1B/C, AR
6A/B, HBC~494 \citep[][and references
  therein]{pueyo2012,nogueira2023}. The archetypical FU~Ori system,
FU~Orionis itself, is a known binary composed of two stars separated
by 0.5\arcsec \, where the northern component (FU~Ori~N hereafter) is
the optically visible star driving the outburst. \citet{wang2004}
discovered the southern binary component, which according to a
spectroscopical analysis by \cite{beck2012}, is the more massive of
the two (hence FU~Ori~S is the primary member).  Early ALMA
Observations of FU~Ori at 0.5\arcsec \, resolution resolved the binary
system in continuum and revealed complex kinematical features 
\citep{hales2015}. High-resolution ALMA observations of FU~Ori at
0.04\arcsec resolution by \citet{perez2020} resolved each
individual disk, reporting very compact sizes possibly truncated by binary interactions, and kinematical signatures of Keplerian
rotation around the outbursting source (FU~Ori~N). These discoveries
contribute to the ongoing debate on whether the enhanced accretion can
be attributed to close companion interactions. Another perspective
posits perturbation by sub-stellar companions \citep{clarke2005}.

\begin{figure*}[h!]

\includegraphics[width=\textwidth]{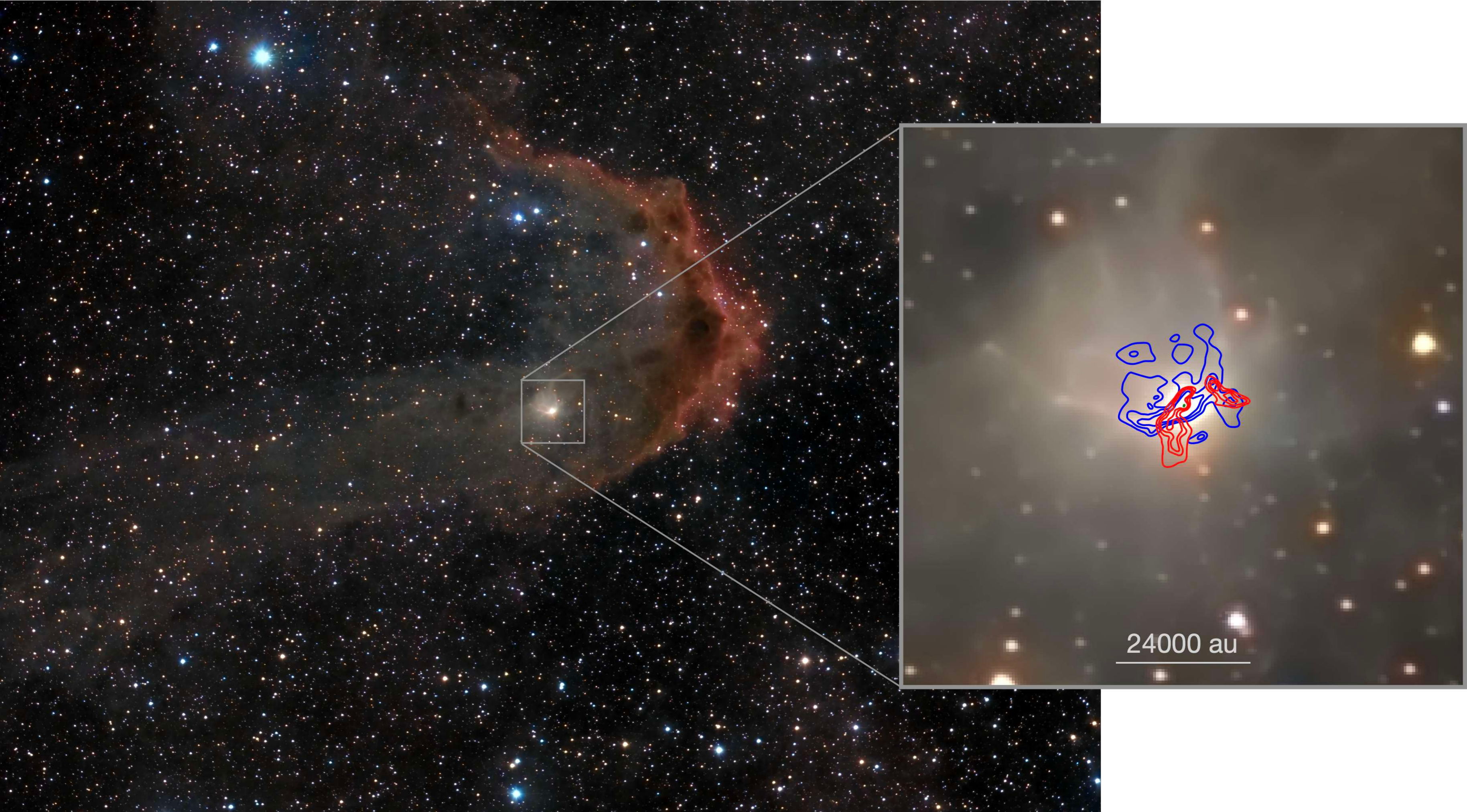}
\caption{{\it{Left}}: Optical {\it{RGB}}  composite image of LBN~878 (the red and brown nebula dominating the field) obtained by astro-photographer Jim Thommes.  FU~Ori (and its reflection nebula) is the bright object located at the center of the image. Inset shows the integrated intensity $^{12}$CO(2--1) maps as traced by the ALMA observations (see Figure~\ref{fig2} for details). Redshifted and blueshifted $^{12}$CO integrated intensity maps of FU~Ori are plotted over optical emission (color scale). The blue-shifted moment 0 map (blue contours) was constructed including emission from 8.0 to 11.5~km~s$^{-1}$ while the red-shifted integrated emission (red contours) includes the emission between 12.7 and 17.5~km~s$^{-1}$. The contour levels are 3, 5, and 8$\sigma$. \label{fig1}}
\end{figure*}

In terms of the evolutionary stage, FUor objects seem to resemble Class~I protostars more than Class~II objects.  Many FUors have Class~I
type SEDs \citep[e.g.,][]{gramajo2014} as well as dust silicate
absorption features, and therefore have been associated with Class~I
protostars \citep{quanz2006}.  Most FUors are surrounded by
reflection nebulae, remnants from their parent cores, some of which are seen to
brighten up during the outburst \citep{herbig1966,Goodrich1987,aspin2003}. FU~Ori itself, located near the Lynds Bright Nebula 878 (LBN~878),
has a fan-shaped reflection nebula that appeared next to the star after the flare, which was not present before \citep{wachmann1954}. Figure~\ref{fig1} a shows large-scale view of the FU~Ori environment, as seen in optical wavelengths. The optical composite image is courtesy of astrophotographer Jim Thommes\footnote{https://jthommes.com/Astro/LBN878.htm}, also known for the discovery of the Thommes Nebula, a reflection nebula that appeared after the outburst of FUor-type object V900~Mon \citep{thommes2011}.

The enormous energy released during the outburst can power winds and
jets which in turn may help disperse the primordial core, a process
that will eventually regulate the final stellar mass \citep{shu1987}.
FUors power prominent winds measured via the presence of P~Cygni
profiles in some specific lines \citep{connelley2018}, from which the
inferred mass loss rates correspond to 10\% of mass accretion rate,
thus several orders of magnitude larger than those of classical T~Tauri stars
\citep{calvet2004}. The observed profiles can be well reproduced by
the standard magneto-centrifugally driven wind model with the addition
of an inner disk wind \citep{milliner2019}.

Early single-dish searches for molecular outflows in FUor objects detected $^{12}$CO(3-2) in many targets, except FU~Ori \citep{evans1994}.
The advent of ALMA  enabled high-resolution interferometric spectral line observations, providing insights into the active circumstellar environments of FUors. These environments are characterized by conspicuous, wide-angle, and slow-moving outflows\citep{ruiz2017a,ruiz2017b,zurlo2017}.

On the other hand, EXors seem to show little outflowing
activity. Outflows have been reported around the ambiguously
classified FUor/EXor sources V1647~Ori and GM~Cha
\citep{cieza2018,principe2018,hales2020}, and only tentatively around
EX~Lup \citep{hales2018,Rigliaco2020}. Although the sample size is
small, the differences in outflow activity between FUors and EXors
suggest that the two types of objects represent different evolutionary
stages, with EXors more evolved than FUors.

There is also observational evidence that most FUor sources are
surrounded by large envelopes that are still transferring material
onto the disk \citep{kospal2017a, feher2017}. It has been proposed
that the imbalance between mass transferred from the envelope to the
disk and the accretion of disk mass onto the star may trigger the
disk instability responsible for the outburst
\citep{bell1994,kospal2017b}. Studying the larger-scale structure of
eruptive sources is thus crucial for understanding the nature of the
FUor/EXor outbursts. 

Asymmetric, non-keplerian gaseous structures are now routinely
discovered by ALMA. Kinematic modeling of these structures reveals
that they can be interpreted as anisotropic infall of material from
the surrounding medium into the circumstellar disk. These structures
end up following elongated patterns for which they have been named
'accretion streamers' \citep[see ][for a
  review]{pineda2023}. Accretion streamers have now been detected
around many Class~0 and Class~I sources and are relevant for improving
our understanding of low-mass star formation. For instance, their
discovery directly challenges the traditional model of axisymmetric
collapse of protostellar cores \citep{shu1977}. The role accretion
streamers play in episodic accretion processes still needs to be
demonstrated; to our knowledge, no accretion streamers have been
discovered around eruptive stars yet.\\


In this work, we present ALMA observations of the surroundings of
FU~Ori at angular scales ranging from 160 to 25000~au at 1.3mm
continuum and CO isotopologes.  Section~\ref{obs} provides details on
the ALMA observations and data reduction, while Section~\ref{results}
presents the primary observational findings and their analysis. These
results are subsequently discussed in Section~\ref{discussion}, and
Section~\ref{conclusion} offers a summary of the study's
findings. Throughout the paper, we assume a distance to FU~Ori of
$408\pm3\,$pc \citep[Gaia DR3;][]{2021AJ....161..147B}.

\section{Observations and Data Reduction}\label{obs}


Observations of FU~Ori (project code 2017.1.00015.S, PI: J. Williams)
were carried out between November 2017 and August 2018 using the ALMA
12-m, 7-m, and Total Power (TP) arrays in order to obtain spatial
coverage from 0.4\arcsec \, to up to a map size of 1\arcmin \,(160 to 25000 au). The
log of the observations, calibrators used, weather conditions, and
array characteristics are shown in Table~\ref{log1}. The full map
required a 27-pointing mosaic for the 12-m array and a 7-pointing
mosaic for the 7-m. The total antenna number for each 12-m, 7-m, and
TP arrays were 45, 11, and 3, respectively. 

The correlators were configured in Frequency Division Mode (FDM) to
observe $^{12}$CO(2--1), $^{13}$CO(2--1), and C$^{18}$O(2--1) in a
single Band 6 spectral setting, providing spectral resolutions of
141~kHz for $^{12}$CO(2--1) and 122.070~kHz for $^{13}$CO(2--1) and
C$^{18}$O(2--1) corresponding to velocity resolutions of 0.184, 0.166
and 0.167~km~s$^{-1}$, respectively. The bandwidth coverage was 142.4,
79.7, and 80.0~km~s$^{-1}$ for $^{12}$CO(2--1), $^{13}$CO(2--1), and
C$^{18}$O(2--1).  A 1.875~GHz-wide spectral window centered at
232.470~GHz was positioned to detect the continuum dust emission.  




\begin{deluxetable*}{lcccccccccccc}
\tablecaption{Summary of ALMA Observations \label{log1}}
\tablewidth{700pt}
\tabletypesize{\scriptsize}
\tablehead{
\colhead{Array} & 
\colhead{Date} &
\colhead{\# Pointings} &
\colhead{UID} & 
\colhead{PWV} &
\colhead{Flux } &
\colhead{Phase} &
\colhead{Baseline} &
\colhead{Elevation} &
\colhead{Time on } &
\colhead{AR} &
\colhead{MRS }
\\
\colhead{} & 
\colhead{} & 
\colhead{} &
\colhead{} & 
\colhead{ (mm)} &
\colhead{Cal.} &
\colhead{Cal.} &
\colhead{Max-Min (m)} &
\colhead{(deg)} &
\colhead{Source (min)}&
\colhead{(\arcsec)} & 
\colhead{(\arcsec)} 
}
\startdata
12m & 2018-01-11 & 27 &-& 2.5 &  J0510+1800 & J0532+0732    & 2516.9-15.1 & 50 & 13.6 & 0.2&2.7 \\
7m & 2017-11-02  & 7  &-& 0.7 &  J0522-3627 & J0532+0732 & 48.9-8.8       & 65 & 37.5 & 4.9&27.4\\
 7m & 2017-11-02 & 7  &-& 0.7 &  J0522-3627 & J0547+1223 & 48.9-8.8       & 55 & 37.6 & 4.9&27.4\\
 TP & 2018-07-11 &-   &-& 2.2 &       -     &     -      & -              & 54 & 38.1 &    &\\
 TP & 2018-07-15 &-   &-& 1.5 &       -     &     -      & -              & 53 & 38.1 &    &\\
 TP & 2018-08-09 &-   &-& 0.6 &       -     &     -      & -              & 54 & 38.1 &    &\\
 TP & 2018-08-09 &-   &-& 0.5 &       -     &     -      & -              & 54 & 38.1 &    &\\
 \hline             
 \enddata
\end{deluxetable*}

All data were calibrated by the ALMA staff using the ALMA Science
Pipeline (version 40896 of Pipeline-CASA51-P2-B) in
CASA~5.1.1\footnote{\url{http://casa.nrao.edu/}}
\citep{mcmullin2007,casa2022}. The calibration process includes offline Water
Vapor Radiometer (WVR) calibration, system temperature correction, as
well as bandpass, phase, and amplitude calibrations. Online flagging
and nominal flagging such as shadowed antennas and band edges were
applied during calibration.  Continuum subtraction in the visibility
domain was performed on the interferometric data prior to imaging.




Imaging of the continuum and molecular emission lines was performed
using the {\sc tCLEAN} task in CASA 6.1.  The TP observations were
converted into visibilities using the task {\sc TP2VIS}
\citep{koda2019}, and then input into {\sc tCLEAN} along with the
interferometric data.

The 12-m array data was self-calibrated using the standalone version
of the automated self-calibration module (Tobin et al., in prep.). The
self-calibration process was performed separately for each
observation, with one iteration of phase-only self-calibration. The
improvements in signal-to-noise (S/N) after self-calibration ranged
between 10-20\%. The self-calibration tables derived from the
continuum data were applied to the FDM spectral windows before imaging
the CO lines. The continuum and spectral line data were imaged using
the {\sc tCLEAN} task interactively in CASA (using \ahg{a} Hogbom deconvolver, creating
manually-defined CLEAN masks, and using
natural weighting to maximize sensitivity). Continuum subtraction was performed before
imaging each molecular line using the task {\sc UVCONTSUB}. During
{\sc clean}ing of the spectral lines, the spectral axis was binned {\it{ on-the-fly}} to
0.25~km~s$^{-1}$ to increase the S/N. The resulting spatial
resolution for the combined 12m+7m+TP dataset corresponds to a
synthesized beam size of 0.4$\arcsec$ $\times$ 0.3$\arcsec$ (PA =
41.6$^{\circ}$).

Integrated intensity (moment 0) and intensity-weighted mean velocity
(moment 1) maps for the spectral line data were produced with CASA
task {\sc IMMOMENTS}. The moment 0 images were produced by integrating
the signal in channels with emission above 3$\sigma$. There is a strong
cloud contamination close to the system's velocity \citep[11.75~km~s$^{-1}$;][]{hales2015}, particularly
strong in the 7m+TP data; thus, for display purposes, we created
separate moment 0 maps for the red-shifted and blue-shifted portions
of the observed large-scale emission.

\section{Results}\label{results}

We detect $^{12}$CO, $^{13}$CO, and C$^{18}$O in a variety of angular
scales.  The source large-scale structure mapped by the 7-m+TP arrays
dominates the final image, overwhelming the structures traced by the
12-m array. For this reason, we also produced continuum and spectral
line cubes using the 12-m data only. This approach allows for a more
detailed examination of the specific characteristics and features
contained in each dataset.

Figure~\ref{fig1} (inset) shows the integrated intensity (moment 0)
map for $^{12}$CO, and Figure~\ref{fig2} shows the integrated
intensity maps for all three CO isotopologues at the spatial scales
traced by the 7-m+TP arrays. The angular resolution of the combined
12+7m+TP image is 0.4\arcsec$\times$0.3\arcsec. However, the maps in
Figure~\ref{fig2} have been smoothed to 2.5\arcsec \, to enhance the
sensitivity to large-scale structures (the nominal resolution of
7m-only data is 4.9 \arcsec \, whereas the angular resolution of the
12m-only data is 0.2 \arcsec). Figure~\ref{fig3} displays the
$^{12}$CO and $^{13}$CO moment 1 maps obtained by imaging the
12m-array data only. C$^{18}$O was not detected in the 12-m array
data.



\subsection{Wide, slow-moving outflows}\label{7m}

\begin{figure*}
\gridline{\fig{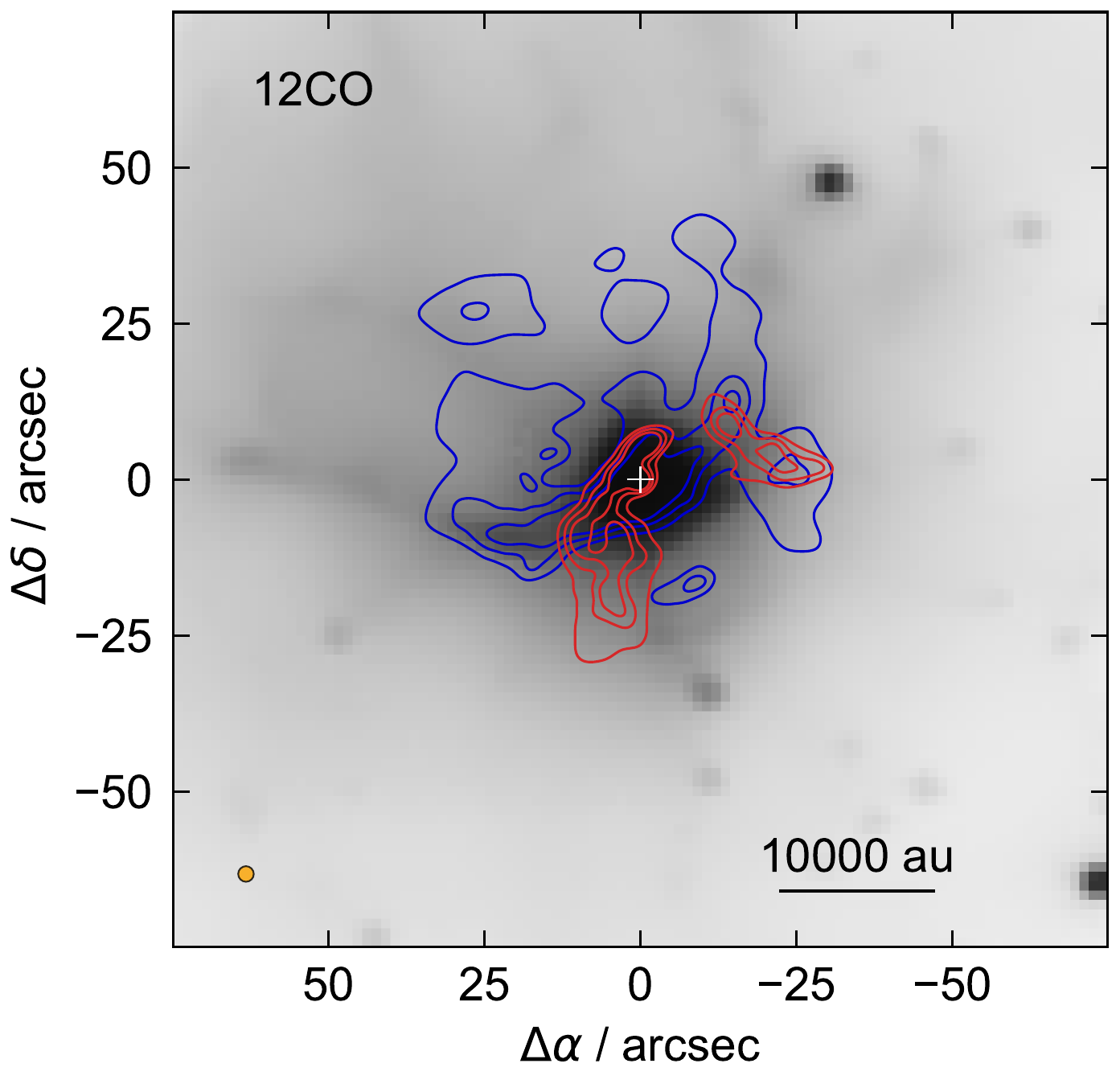}{0.3\textwidth}{(a) }
          \fig{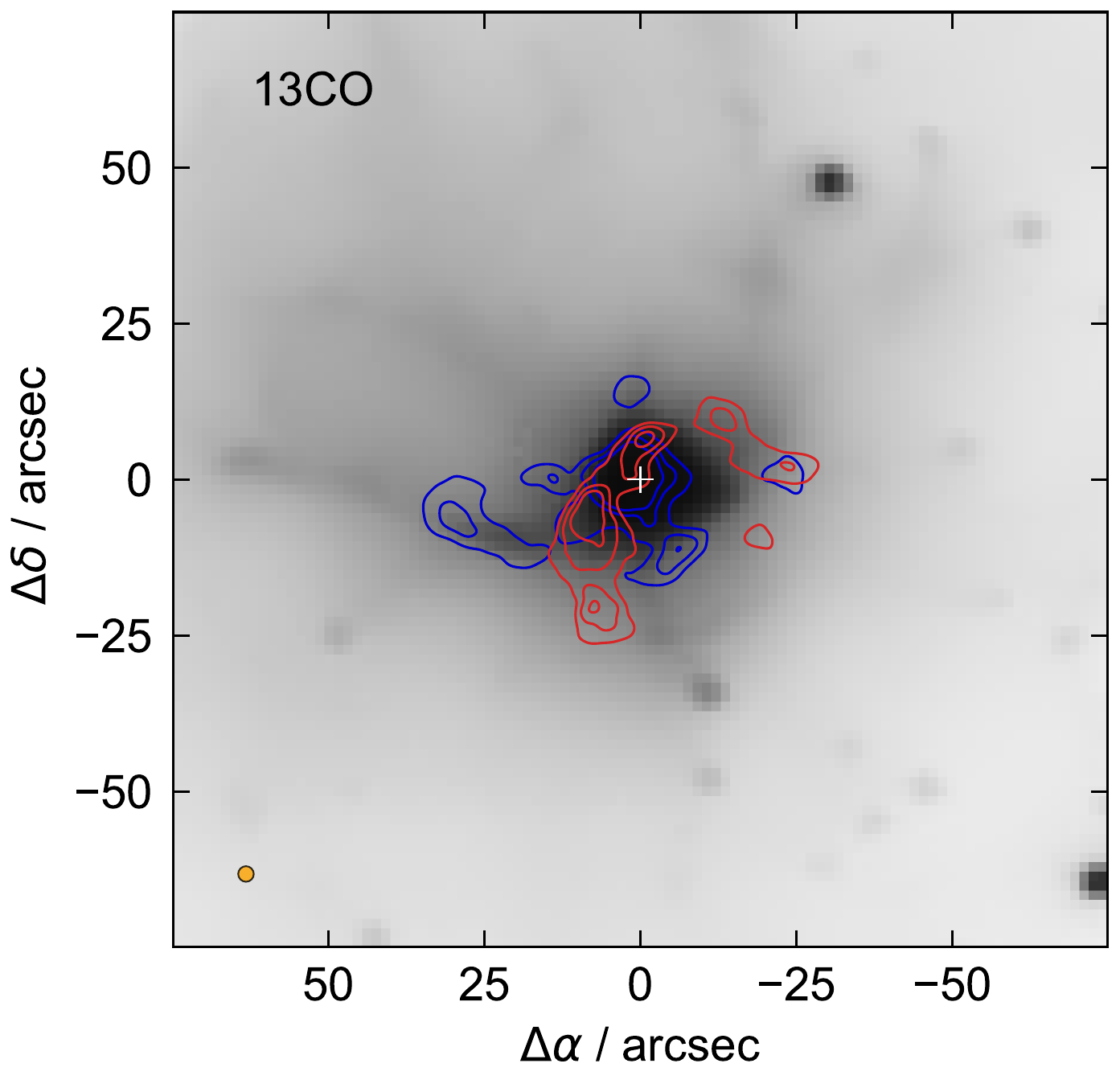}{0.3\textwidth}{(b) }
          \fig{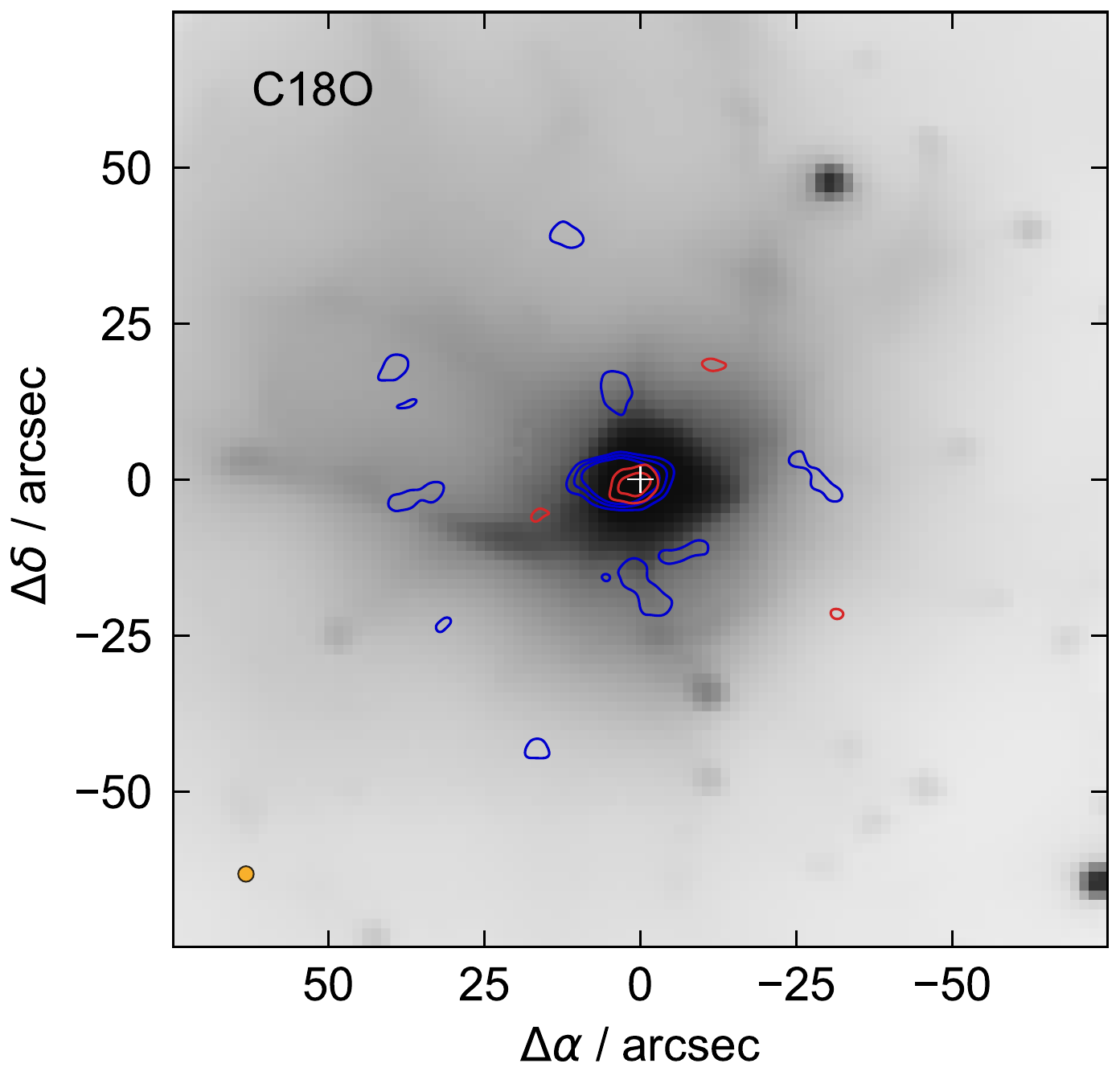}{0.3\textwidth}{(b) }
          }
\caption{Integrated intensity in FU~Ori maps for $^{12}$CO, $^{13}$CO
  and C$^{18}$O. The blue-shifted moment 0 maps (blue contours) were
  constructed including emission from 8.0 to 11.5~km~s$^{-1}$ while
  the red-shifted images (red contours) include the 12.7 to
  17.5~km~s$^{-1}$. Grayscale shows the optical image from Figure~\ref{fig1}. The white cross at the
   center of the image shows the position of FU~ori~N. The angular resolution of the
  images (2.5\arcsec$\times$2.5\arcsec) is shown as a black circle in the lower left.\label{fig2}}
\end{figure*}

The ALMA observations reveal the presence of large-scale, wide-angle
bipolar outflows for the first time around the class prototype FU
Ori. The blue-shifted portion of the $^{12}$CO emission traces the
northeastern part of the well-known reflection nebula.  The $^{13}$CO
emission partially coincides with the $^{12}$CO emission
(Figure~\ref{fig2}), whereas the C$^{18}$O is more compact. The
$^{12}$CO and $^{13}$CO emission have an extension of approximately
20000~au. The line-of-sight velocity of the outflows is low, typically
around $\sim$1-2~km~s$^{-1}$. The wide-angle, slow-moving bipolar
outflow traced by $^{12}$CO emission is reminiscent of the emission
detected towards other FUor objects: V883~Ori \citep{ruiz2017a},
HBC~494 \citep{ruiz2017a}, V2775~Ori \citep{zurlo2017}, V1647~Ori
\citep{principe2018}, V900~Mon \citep{takami2019}, and GM~Cha
\citep{hales2020} as well as in low-mass Class I young stellar objects
\citep[e.g. ][]{arce2006,gonzalesinprep}. This slow outflow is also
comparable to those seen towards First Hydrostatic Core candidates
\citep[e.g. ][]{Pineda2011,Maureira2020}.

\subsection{Streamer in FU~Ori}\label{12m}

\begin{figure*}

\includegraphics[width=0.49\textwidth]{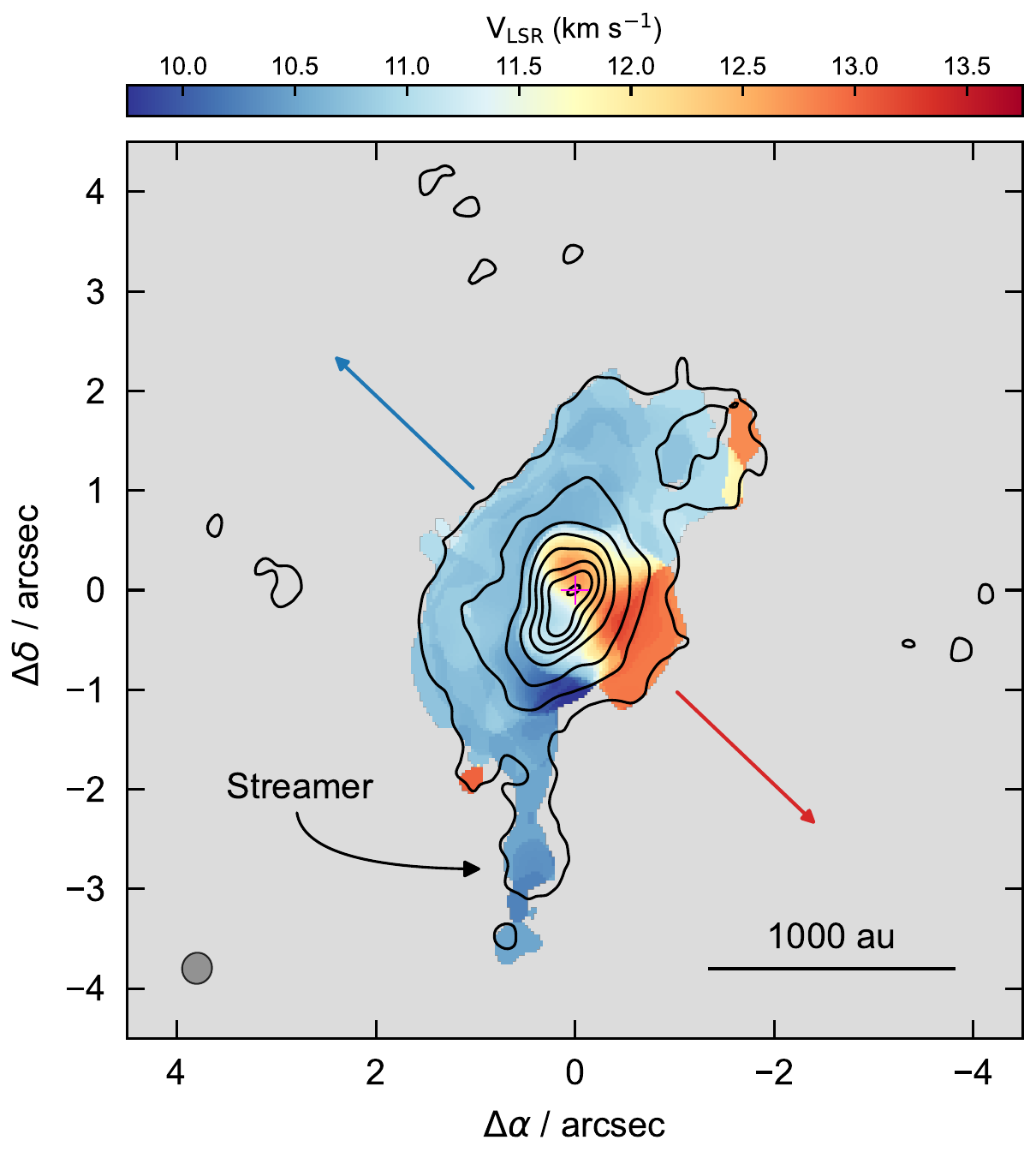}
\hfill
\includegraphics[width=0.49\textwidth]{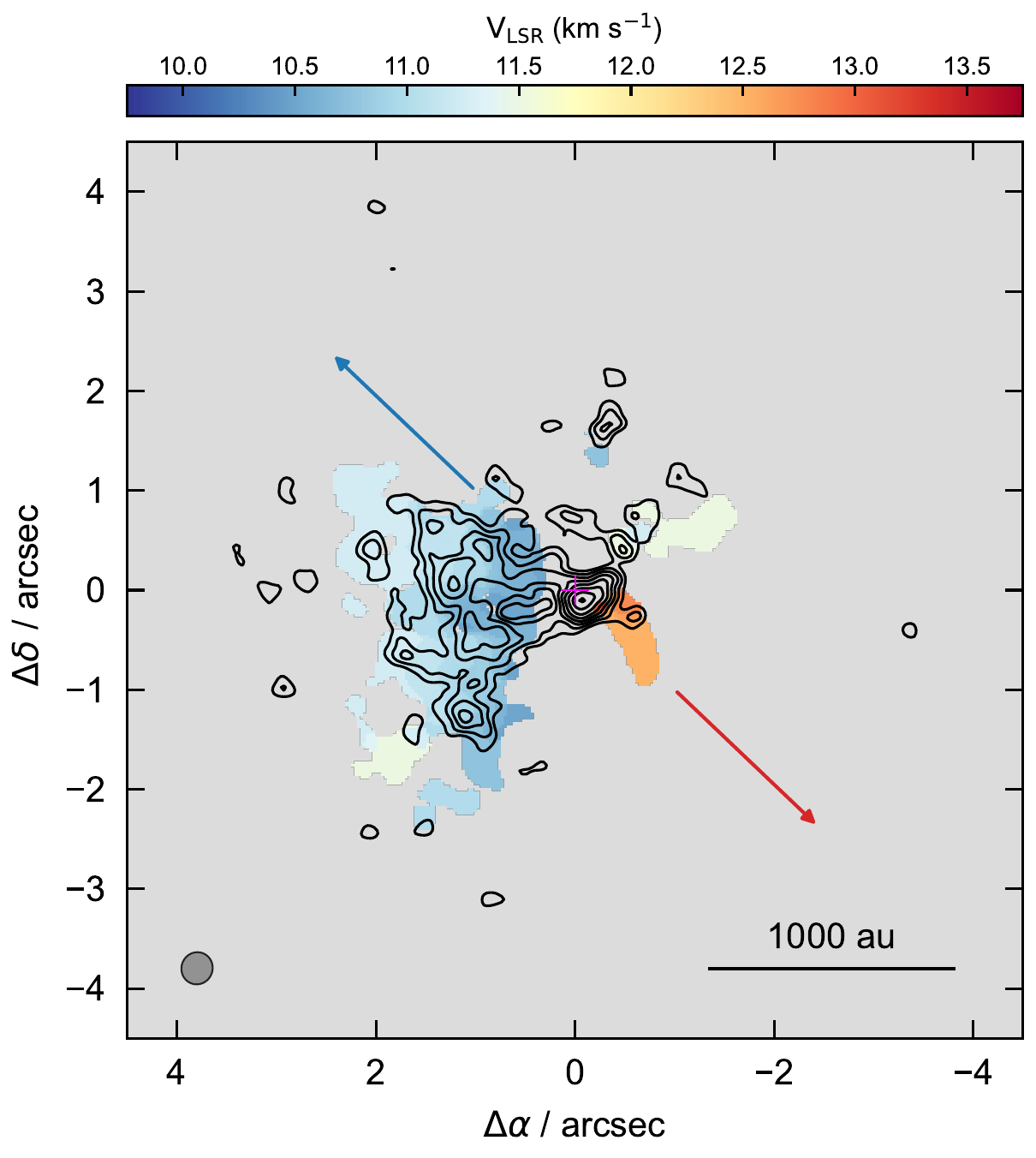}
\caption{Zoomed-in view of FU~Ori in $^{12}$CO and $^{13}$CO emission generated imaging the 12m-array data only. Both maps show the velocity centroids (moment 1) in {\tt jet} color scale ranging from 9.5 to 13.5~km~s$^{-1}$ (centered on the systemic velocity of 11.75~km~s$^{-1}$). The moment zero maps are shown in contours ranging from 3 times the zeroeth moment RMS to the peak. Contours range from 39 to 707~mJy\,beam$^{-1}$~km~s$^{-1}$, and 26 to 78~mJy\,beam$^{-1}$~km~s$^{-1}$, for $^{12}$CO and $^{13}$CO, respectively. The white cross shows the position of the continuum maximum which we take as a proxy for the position of FU~Ori~N. The ellipse on the lower left represents the synthesized beam. The blue and red arrows show the expected direction of the large-scale outflow assuming it is perpendicular to the disk PA of 133.6$^\circ$ measured from the long baseline data in \citet{perez2020}. \label{fig3}} 
\end{figure*}

Figure~\ref{fig3} shows the zoomed-in view of the $^{12}$CO(2-1) and
$^{13}$CO(2-1) integrated velocity maps (moment 1) of FU~Ori, as revealed by the 12-m array data only. The
intermediate-scale emission traced by the 12-m Array shows
complex kinematics. The $^{12}$CO(2-1) map is similar to the Cycle~0
$^{12}$CO(3-2) map presented in \citet{hales2015}, yet significantly
more extended. The elongated feature extending south of the
central objects at velocities 10-11~km~s$^{-1}$, also observed in the
Cycle~0 $^{12}$CO(3-2) channel maps, is now confirmed to be real by our deeper,
higher-resolution $^{12}$CO(2-1) data. Linear structures have been
observed in scattered light around FUor type objects
\citep{liu2018,takami2018}, however, this is the first time such a structure is reported in molecular line emission. Furthermore, the feature we
observe in $^{12}$CO does not coincide with any of the linear features
reported in the near-IR. The elongated feature has neither connection to the
larger-scale molecular outflow nor to the inner disk rotation \citep{perez2020}, and
is more similar to accretion streamers recently reported around young
stellar objects \citep{pineda2020,alves2019,valdivia-mena2022,Gupta2023,pineda2023}.

\subsection{Outflow properties}\label{outflow}

Swept-up cavities in the envelope of young stellar objects play a significant role in understanding the complex processes occurring during their early stages of formation \citep[][and references therein]{arce2007}. These cavities are created by the powerful outflows and jets emitted from protostars, which interact with their surrounding molecular clouds. As the outflows propagate through the dense material, they sweep away and evacuate cavities, shaping the surrounding environment. These cavities serve as valuable indicators of the energetic feedback mechanisms associated with star formation. By studying the size, shape, and kinematics of these swept-out cavities, researchers can gain insights into the dynamics of
protostellar outflows, the interaction between the outflows and their surroundings, and the impact of these processes on the overall evolution of young stellar objects.\\

\begin{figure}[h!]
\centering
\includegraphics[width=\columnwidth]{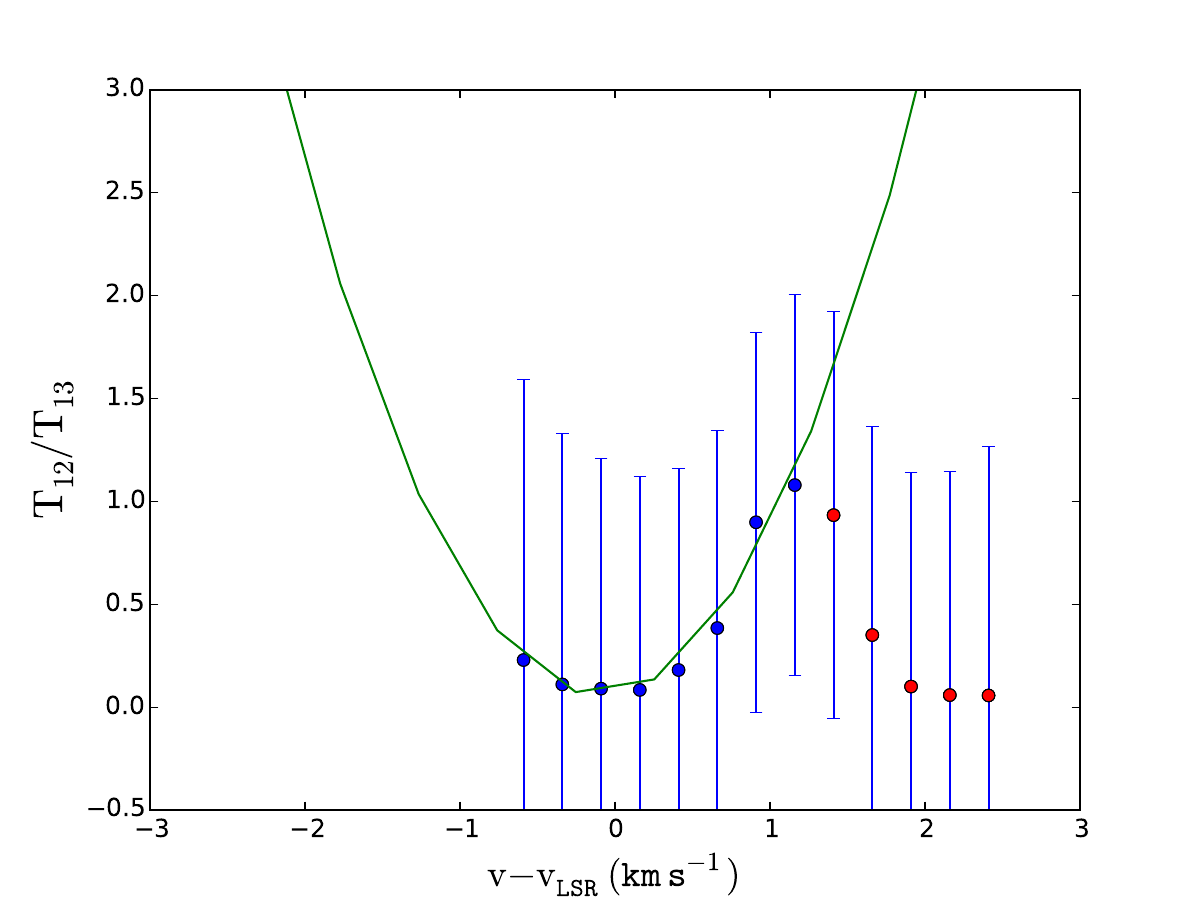}
\caption{{Relationship between the brightness temperatures of
        $^{12}$CO (T$_{12}$) and $^{13}$CO (T$_{13}$) presented as a function of
        velocity relative to the systemic velocity. The plotted data
        includes representing the weighted mean values (blue solid
        dots), accompanied by error bars indicating the corresponding
        weighted standard deviations in each channel. The red solid
        dots depict the weighted mean values that were not employed
        during the fitting since the emission is optically thin in
        those channels and, therefore, no correction is needed. To
        represent the data, a green solid line is used to illustrate
        the best-fit second-order polynomial. \label{fig4}  }}
        
\end{figure}

Assuming both the $^{12}$CO and $^{13}$CO lines trace the bipolar
cavity, we use the $^{12}$CO and $^{13}$CO emissions to derive
estimates of the mass of the outflow and its kinematic properties in
the traditional manner \citep[e.g.;
][]{cabrit1990,Dunham2014}. Following the process described in
\citet{ruiz2017a}, we estimate the outflow quantities from the blue-
and red-shifted emissions separately. The optically-thin tracer
$^{13}$CO is used to correct for the optical depth effects in
$^{12}$CO by computing the ratio of the brightness temperatures of
each line from all the channels with detections above 5$\sigma$. To
apply the correction factor to data with only $^{12}$CO detection, we
extrapolate values from a parabola fitted to the weighted mean
values. Figure~\ref{fig4} shows the best fit parabola (solid green line). The
derived outflow properties are shown in Table~\ref{table2}.



\begin{table}
\begin{center}

\caption {Mass, Momentum, Luminosity, and Kinetic Energy of the Outflow} \label{table2}
\begin{tabular}{lcc}
\hline
 Blue Lobe    &    & \\
\hline
 & 20 (K)                         & 50 (K)         \\
\hline
 Mass    (M$_{\odot}$)   & 1.37              & 2.08          \\
 Mass loss (M$_{\odot}$~yr$^{-1}$) & 3.6$\times 10^{-4}$  & 5.5$\times 10^{-4}$ \\
 Momentum   (M$_{\odot}$~km~s$^{-1}$) & 1.16      & 1.76    \\
 Energy (erg)     & 1.02$\times 10^{43}$           & 1.55$\times 10^{43}$            \\
 Luminosity (L$_{\odot}$) & 0.02            & 0.03         \\
 \hline \\
  \hline
 Red Lobe    &    & \\
 \hline
 & 20 (K)                         & 50 (K)         \\
 \hline
 Mass     (M$_{\odot}$)  & 0.019       & 0.029       \\
 Mass loss (M$_{\odot}$~yr$^{-1}$) & 5.1 $\times 10^{-6}$ & 7.7$\times 10^{-6}$ \\
 Momentum (M$_{\odot}$~km~s$^{-1}$)  & 0.014  & 0.022 \\
 Energy (erg)     & 1.08$\times 10^{41}$          & 1.64$ \times 10^{41}$           \\
 Luminosity (L$_{\odot}$) & 2.3$\times 10^{-4}$       & 3.5 $\times 10^{-4}$        \\
\hline 
\end{tabular}
\end{center}

\end{table}


\subsection{Streamer Model} \label{streamer_model}


To ascertain the infalling nature of the observed streamer, we fitted
$^{12}$CO(2--1) streamer emission (see Figure~\ref{fig3}) with
analytical equations of infalling trajectories given by
\citet{mendoza2009}. For this fitting, we use TIPSY \citep[Trajectory of
Infalling Particles in Streamers around Young stars; ][]{Gupta2024} which simultaneously fits the morphology and velocity profile of
the streamer observations.  TIPSY requires prior estimation of the mass of
the central object and systemic line-of-sight velocity of the center of
mass, which were taken to be 1.8~M$_{\odot}$ and 11.75~km~s$^{-1}$, respectively, since the masses of FU~Ori~N and FU~Ori~S are 0.6 and 1.2~M$_{\odot}$ respectively. \citep{perez2020,beck2012}. 
\cite{perez2020} inferred the disk dust masses for FU~Ori~N and FU Ori~S to be $22$~M$_\oplus$ and 8~M$_\oplus$, respectively.
Assuming the typical gas-to-dust ratio of 100, the total masses of the disks are expected to be 
$\sim7$~M$_{\text{jupiter}}$ (or $\sim6.6\times10^{-3}$~M$_{\odot}$) and $\sim3$~M$_{\text{jupiter}}$ (or $\sim2.8\times10^{-3}$~M$_{\odot}$), respectively, which are negligible for the streamer model calculations.



\begin{figure*}
\centering
\includegraphics[height=0.4\textwidth]{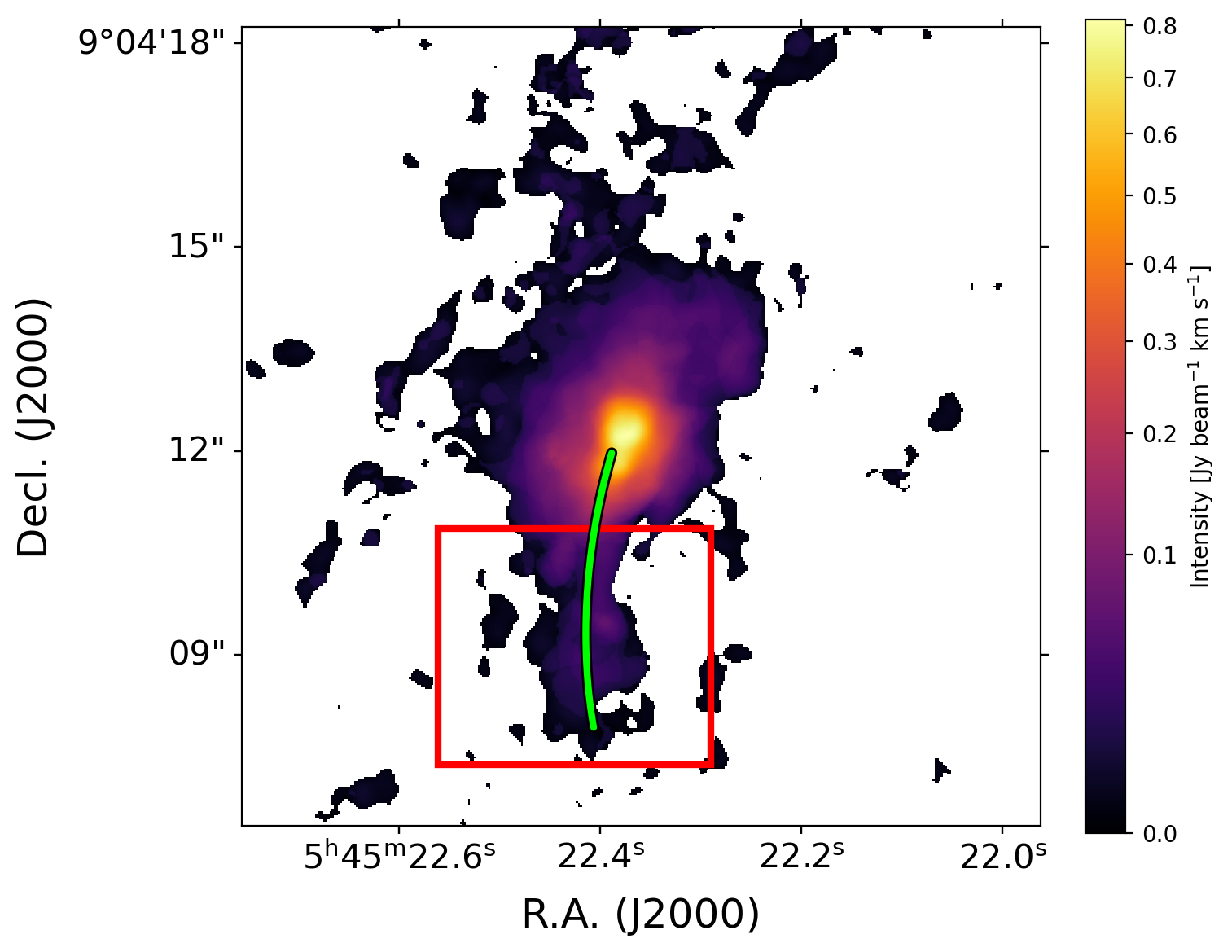} 
\includegraphics[height=0.4\textwidth]{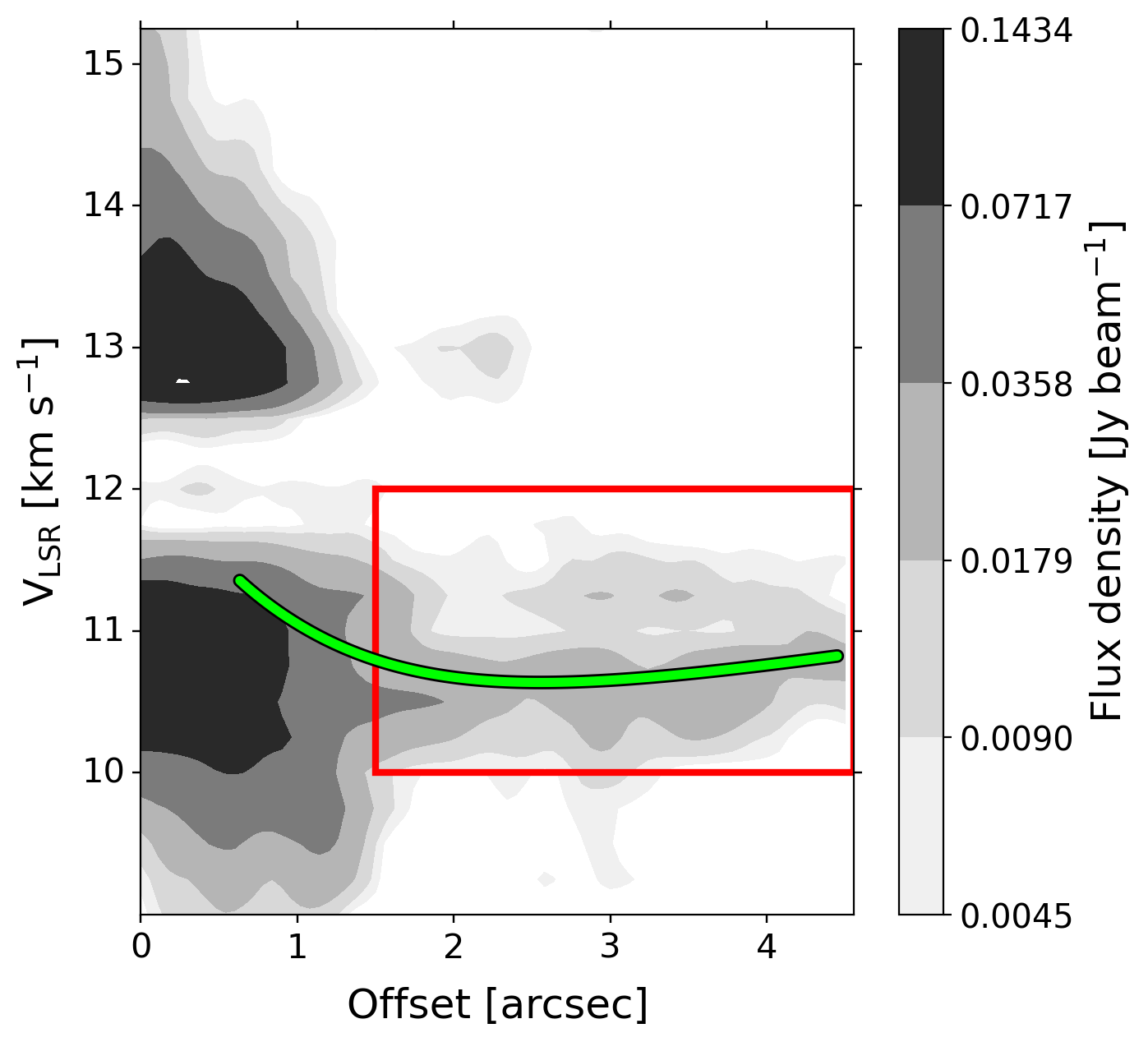}
\caption{The best fit infalling trajectory from TIPSY (green curved line) overplotted on the $^{12}$CO integrated intensity map (left panel) and position-velocity (PV) diagram along the streamer (right panel). In the PV diagram, the outermost contours represent the noise level of the data and each subsequent contour denote an increment by a factor of two. The red rectangles in both panels represent the initial  R.A., Decl., and {V}$_{\rm LSR}$ limits given to TIPSY to extract the sub-cube of the streamer emission. A clustering algorithm does the rest of fine-tuning of the extraction \citep[see ][ for details]{Gupta2024}.\label{fig:tipsy_fit} } 
\end{figure*}

The fitting results suggest that the morphology and the velocity
profile of observed streamer emission can be well-represented as a
trail of infalling gas, as shown in Figure \ref{fig:tipsy_fit}.  Using
the distribution of trajectories which can fit the observed streamer,
infalling timescale, of the streamer was estimated to be $8537\pm 3317$
year.
Here the infalling timescale is defined as the time taken for infalling material to reach the closest point to the protostars in their trajectory, starting from the farthest point in the observed streamer.
The uncertainties were computed as the standard deviation of infalling timescales, as estimated for the trajectories that could fit the data reasonably well \citep[see Sec. 2.3. in][]{Gupta2024}.
The uncertainty on our trajectory solution is mainly due to lack of curvature in the projected streamer morphology, which is useful in constraining gas velocities along the plane of the sky.
Note that the theoretical trajectories used for the fitting
assume that the only force acting on the particle is gravity from a
point source. Although FU~Ori is a binary system, this assumption is
still valid at the length scale of the streamer which is roughly six times
the binary separation.

The streamer was also observed in $^{12}$CO(3--2) emission \citep{hales2015}, which
allows us to simultaneously estimate temperature and column density
of the gas in the streamer. We estimated the brightness temperature of
infalling gas to be $11.3$~K and $7.1$~K in $^{12}$CO(2--1) and (3--2)
emission, respectively.  
Using the radiative transfer code, RADEX,
brightness temperatures were found to correspond to the gas temperature of
$\sim25$~K and gas column density of $\sim2.5\times10^{16}$~cm$^{-2}$.
This is also in agreement with the non-detection of the streamer in
$^{13}$CO(2--1) emission.  Assuming the typical CO to H$_{2}$ ratio of
$10^{-4}$, the mass of gas in the streamer is then estimated to be
$\sim5.6\times10^{-7}$~M$_{\odot}$.
Although we use the same boundaries to select streamer emission in 2--1 and 3--2 emission maps, the resulting ratios may be affected by the different angular scales traced by the two datasets (which differ by a factor of 2 in terms of angular resolution and maximum recoverable scale), thus adding extra uncertainty to the derived quantities.

Although these estimates do not include parts of streamers beyond the
primary beams of the interferometric observations, they can still be
used to estimate mass infall rates as
$\dot{M}_{inf}=M_{streamer}/T_{inf}$, where $M_{streamer}$ is observed
streamer mass and $T_{inf}$ infall time for the observed streamer. We
find the streamer to be feeding the FU~Ori system at a rate of
$\sim6.6\times10^{-11}$~M$_{\odot}$~yr$^{-1}$. Note that this
value should be treated as an order of magnitude estimate, because of uncertainties on the streamer mass.  A more
reliable mass infall rate estimation will require modeling the total reservoir of bound gas and ideally, in a more optically thin line emission like $^{13}$CO(2--1), which was not detected in our streamer observations. 


\section{Discussion}\label{discussion}

\subsection{Molecular Outflow}

Our new ALMA observations detect a wide-angle outflow emerging from
the FU~Ori binary system for the first time. The northern component of
the binary system, FU~Ori~N, is the outbursting source. Thus it is
natural to assume that FU~Ori~N was the source driving the
outflows. Prior searches for molecular outflows around FUors, mainly using single-dish telescopes \citep{evans1994}, reported outflowing
material from many FUors but failed to detect flows emerging from the
FUor class prototype. These non-detections instigated the belief that
there were no molecular outflows around the FU~Ori  system. Our discovery ends the
mystery by clearly demonstrating the presence of a molecular outflow from FU~Ori itself.

The main characteristics of the outflow discovered around FU~Ori are
its wide angle and low velocity.  The blue-shifted lobe extends
approximately 10000~au in each direction.  We compute the dynamical
timescale of the outflow by dividing the extension of the blue-shifted
lobe by the maximum gas velocity (not considering correction for the
system's inclination).  We obtain a Kinematic Age of 3800~yr, which is
significantly larger than the typical duration of FUor outbursts
($<10^2$~yr).   This implies that the current outburst is not responsible for the observed outflows, similar to what it was concluded for V883~Ori \citep{ruiz2017b}.

The blue-shifted lobe has a projected opening angle of
$\sim$140-150$^{\circ}$ . Wide-angle outflows have been reported
around other FUors, such as HBC~494, V883~Ori, and GM~Cha, as well as
in Class~I objects such as RNO~129
\citep[e.g.;][]{ruiz2017a,ruiz2017b,hales2020,arce2006}. These
outflows are wider than those of other FUor observed with ALMA such as
V346~Nor, Haro~5a~IRS, V1647~Ori, V2775~Ori and V900~Mon
\citep{kospal2017b,kospal2018,principe2018,zurlo2017,takami2019}. \citet{arce2006}
reported a trend in which outflow opening angles increase with the age
of the protostars. A possible explanation for the widening of outflows
with age is that in the early stages, the main observed outflow
structures are highly collimated, fast-moving jets that begin to
pierce the envelope. As most of the envelope material along the jet
axis is cleared out, the prevailing structure in the molecular outflow
will be primarily shaped by the gas carried along by the wide-angle
portion of the wind. Alternatively, the parabolic morphology of FU~Ori's wide opening molecular outflow is consistent with the shell disk wind model
\citep{lee2000,arce2007,devalon2022}.

In this context, the FU~Ori system is more evolved than other Class~I
protostars that are still heavily embedded in their surrounding
envelope and have narrower cavity opening angles. This interpretation is consistent with the fact that FU~Ori
is optically visible because most of the surrounding material has been
accreted or swept up by  mass dissipation processes
(e.g. primordial high-velocity jets, disk winds, disk+star accretion). The $^{12}$CO and
$^{13}$CO outflows detected by ALMA coincide with the wide-angle
conical reflection nebula seen in optical wavelengths
(Figure~\ref{fig1}), suggesting that the dissipative processes acting
in FU~Ori effectively disperse gas and dust from the primordial
envelope. Nevertheless, there is still a significant amount of
cloud/envelope material, evidenced by the substantial contamination in
systemic velocities (Figures~\ref{12COchan}-\ref{13co_fuor_c435}) and  the
centrally peaked C$^{18}$O emission traced by the 7m+TP array
observations (Figure~\ref{fig2}), likely associated with the remnant of a primordial envelope.

The projected velocity of the outflows is small, with a characteristic
velocity of $\sim$1-2~km~s$^{-1}$. Using relative velocities of 1~km~s$^{-1}$ and a distance from the center of 10000~au, we find the outflow material to be unbound. The low outflow velocity of the
$^{12}$CO and $^{13}$CO outflows is much smaller than the typical
10-40~km~s$^{-1}$ observed around other FUors \cite[][ and references
  therein]{ruiz2017b} but is similar to that of V883~Ori \citep[0.65~km~s$^{-1}$][]{ruiz2017b}. \citet{perez2020} estimated an inclination 37$^{\circ}$
degrees for the dust disk around FU~Ori~N. Thus, assuming that the
outflow's direction coincides with the disk's angular momentum, we
discard that the low velocity observed is due to inclination
effects. \citet{ruiz2017b} proposed that in the case of V883 Ori, the low
outflow velocities can be explained if the outflow we see is very old,
a remnant of older fast-moving outflows that have now ceased or
emerging from a quiescent disk without the creation of high-velocity
components, similar to what we observe in the FU~Ori system. FU~Ori is known to have deep, broad absorption P~Cygni profiles that can be described by disk wind models capable of sustaining mass-loss rates of the order of 10\% the accretion rate \citep[i.e. mass loss due to wind of $\sim 10^{-5}$ to $10^{-6}$~M$_{\odot}$~yr$^{-1}$;][]{milliner2019}. 

The properties of the outflow listed in Table~\ref{table2} can be readily
compared with those of V883~Ori and HBC~494, as they were derived
using the same method. The
estimated mass for the blue-shifted lobe is two orders of magnitude
larger than the red-shifted lobe. We cannot confirm whether this
difference is real or possibly caused by the absorption of
cloud/envelope material in the line of sight. Nevertheless, the mass
estimation of the blue-shifted lobe is comparable to that of V883~Ori
and one order of magnitude larger than that of HBC~494. The mass loss,
momentum, energy, and luminosity are also similar to those of V883~Ori
yet still consistent with the values detected around other young
protostars \citep[e.g.][]{arce2006}.



\subsection{Accretion Streamer}

As discussed in Section \ref{streamer_model}, the $^{12}$CO streamer
(Figure \ref{fig:tipsy_fit}) seems to be feeding mass to the protostellar
system at a rate of $\sim6.6\times10^{-11}$~M$_{\odot}$~yr$^{-1}$ (or $\sim0.07$~M$_{\text{jupiter}}$~Myr$^{-1}$). 
As mentioned in Section \ref{streamer_model}, the disk masses for FU~Ori~N and FU Ori~S are $\sim6.6\times10^{-3}$~M$_{\odot}$ and $\sim2.8\times10^{-3}$~M$_{\odot}$, respectively.
This would suggest that the observed streamer will require $\sim100$~Myr to replenish disks masses, which is at least an order of magnitude greater than the typical disk lifetimes.

The estimated mass infall rate in the streamer around FU~Ori is lower than those detected in streamers around other Class~I sources \citep[$\sim10^{-6}$~M$_{\odot}\,$yr$^{-1}$, e.g.,][]{valdivia-mena2022,pineda2023}. 
The streamer was not  detected in C$^{18}$O, consistent with it not being as massive as the ones detected around other Class~I protostars. There is also no clear detection of streamer in $^{13}$CO emission, although there is still some hint of stream-like emission in some $^{13}$CO channel (channels 11--11.25 km~s$^{-1}$, see Figure~\ref{13co_fuor_c435}), just adjacent to $^{12}$CO streamer emission (mostly in channels 10.25--10.75 km~s$^{-1}$, see Figure~\ref{12co_fuor_c435}). Therefore, some of the $^{12}$CO streamer emission may be hidden due to high optical depth effects. This may suggest that we are underestimating streamer mass, although not by orders of magnitude as we do not detect the streamer in $^{13}$CO or C$^{18}$O emission.

The streamer needs to be more massive to sustain FU~Ori's outburst accretion rates (by several orders of magnitude). The estimated streamer mass infall rate is not even sufficiently massive to sustain quiescent stellar accretion rates. 
Despite not being massive enough to sustain the outburst accretion rate, the streamer can deliver material to the disk and trigger disk instabilities, which can further lead to accretion. 
Anisotropic infall, cloudlet capture events, inhomogeneous delivery of material, and building-up of material around dust traps can all lead to the disk instabilities that could trigger accretion outbursts \citep{vorobyov2010,dullemond2019, kuznetsova2022,hanawa2024}.  
It still needs to be determined if the low observed mass infall rate is enough to trigger disk instabilities, for which detailed modeling would be required.

It is also possible that the streamer we observe now is not directly connected to the current outburst but instead is a remnant of a more massive streamer that could have played a role in feeding  mass to the disk in the past, but now has already relinquished most of its mass into the protostellar system.  In this scenario, the streamer we are observing is the smoking gun for previous, vigorous episodes of cloud accretion into the disk that may have led the disk to become unstable and trigger  FU~Ori's outburst.

There may also be more than one streamer interacting with the binary system. For instance, the high-resolution ALMA observations of FU~Ori from \citet{perez2020} detected CO emission that delineates the position of the brightest feature in scattered light: a prominent arc to the northeast. This arc could represent a signpost of active accretion into the system or could be light reflected from material at the base of the outflow (Zurlo et al. in preparation), similar to features recently reported in other outbursting systems \citep{weber2023}.

\section{Conclusions}\label{conclusion}

The current study presents ALMA observations of the FU~Ori outbursting
system covering spatial scales from 160 to 25,000 astronomical units.

The remarkable sensitivity to a wide range of spatial scales unveil a
wide, slow-moving molecular outflow. To obtain the first determination
of the outflow properties, including mass, momentum, and kinetic
energy, we have employed $^{13}$CO(2--1) data to correct for optical
depth effects.

A thorough analysis of the higher-resolution interferometric data
reveals intricate gas kinematics, likely resulting from a confluence
of ongoing physical processes, including accretion, rotation,
ejection, and spreading of material due to binary interaction

The new observations corroborate the existence of an elongated
$^{12}$CO feature that can be modeled consistently as an accretion
streamer. The streamer analysis suggests that the stream cannot
sustain the enhanced accretion rate. Nevertheless, the observed
streamer may be feeding the disk, helping to produce conditions of
instability that can give rise to the outburst. Follow-up
hydro-dynamical modeling is required to ascertain the plausibility of
this scenario. Alternatively, the observed streamer might not be
directly tied to the current outburst but instead be a leftover from a
prior, more massive streamer that previously contributed to the disk
mass, rendering the disk unstable and triggering the FU Ori
outburst. These results demonstrate the value of multi-scale
interferometric observations to enhance our understanding of the
FU~Ori outbursting system and provide new insights into the complex
interplay of physical mechanisms governing the behavior of FUor-type
and the many kinds of outbursting stars.


\software{Common Astronomy Software Applications \citep{mcmullin2007,casa2022}, TIPSY \citep{Gupta2024}, and TP2VIS
\citep{koda2019}}

\section*{Acknowledgments}

We express our gratitude to Jim Thommes for generously providing the optical image of FU~Orionis.
This paper makes use of the following ALMA data:
ADS/JAO.ALMA\#2017.1.00015.S. ALMA is a partnership of ESO
(representing its member states), NSF (USA) and NINS (Japan), together
with NRC (Canada) and NSC and ASIAA (Taiwan), in cooperation with the
Republic of Chile. The Joint ALMA Observatory is operated by ESO,
AUI/NRAO and NAOJ. The National Radio Astronomy Observatory is a
facility of the National Science Foundation operated under a
cooperative agreement by Associated Universities, Inc.
This work acknowledges support from ANID -- Millennium
Science Initiative Program -- Center Code
NCN2021\_080. S.P. acknowledges support from FONDECYT grant 1231663. A.S.M. acknowledges support from ANID / Fondo 2022 ALMA
/ 31220025. 
P.W. acknowledges support from FONDECYT grant 3220399.
J. P. W. acknowledges support from NSF grant AST-2107841.


{}

\appendix

\section{Channel maps}\label{pvdiagchanmaps}

\begin{figure*}
\includegraphics[width=\textwidth]{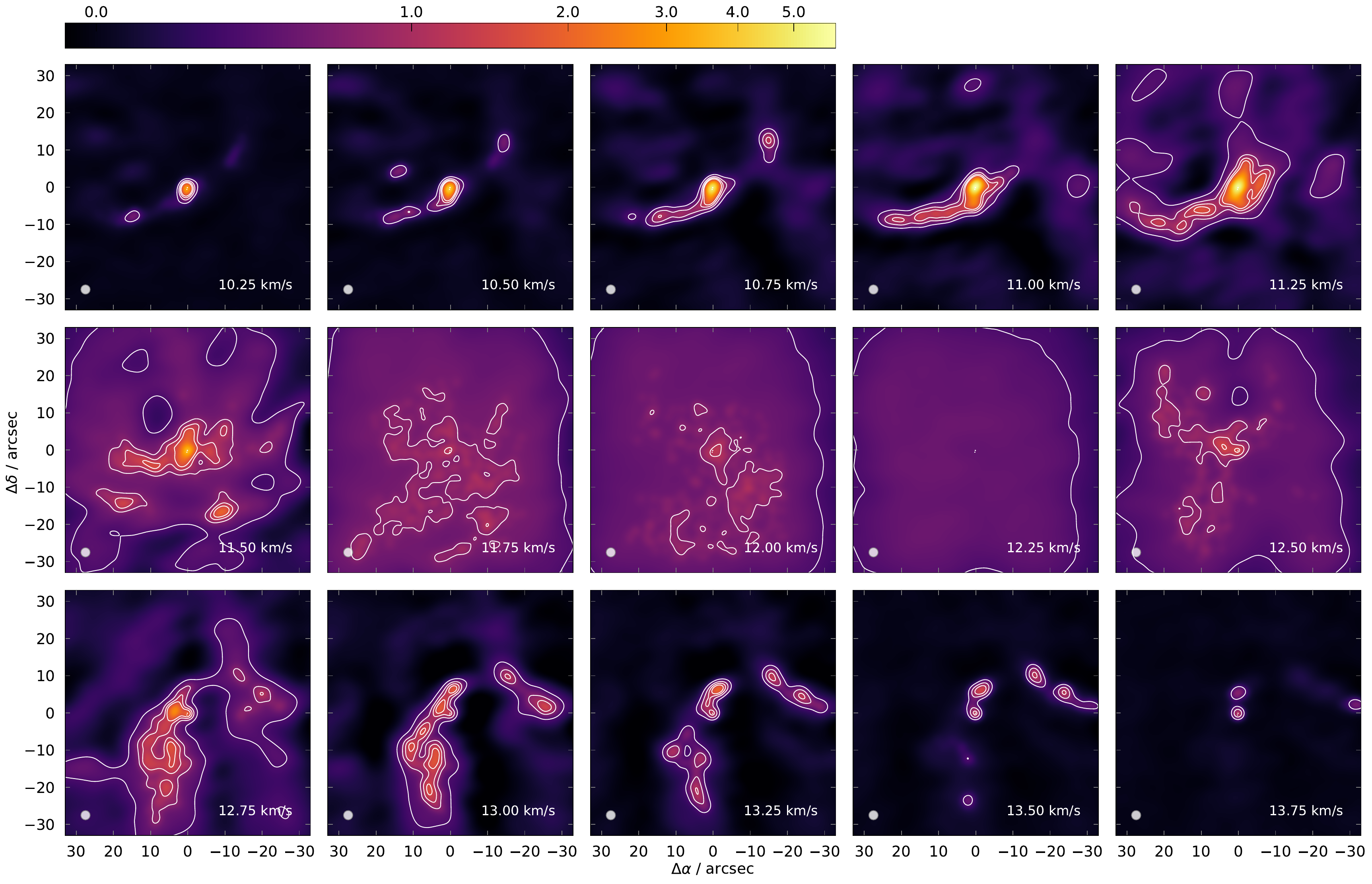}
\caption{$^{12}$CO channel maps towards FU~Ori (7m+TP, smoothed to 2.5'' resolution). The
  velocity of the channels is shown in the Local Standard of Rest
  (LSR) frame, centered at the rest frequency of $^{12}$CO(2--1). The
  data has been binned to a velocity resolution of
  0.25~km~s$^{-1}$. Contours of the 1.3mm continuum emission at 0.3''
  resolution are overlaid (contour levels are 0.5, 1, and 1.8
  mJy\,beam$^{-1}$).\label{12COchan}}
\end{figure*}

\begin{figure*}
\includegraphics[width=\textwidth]{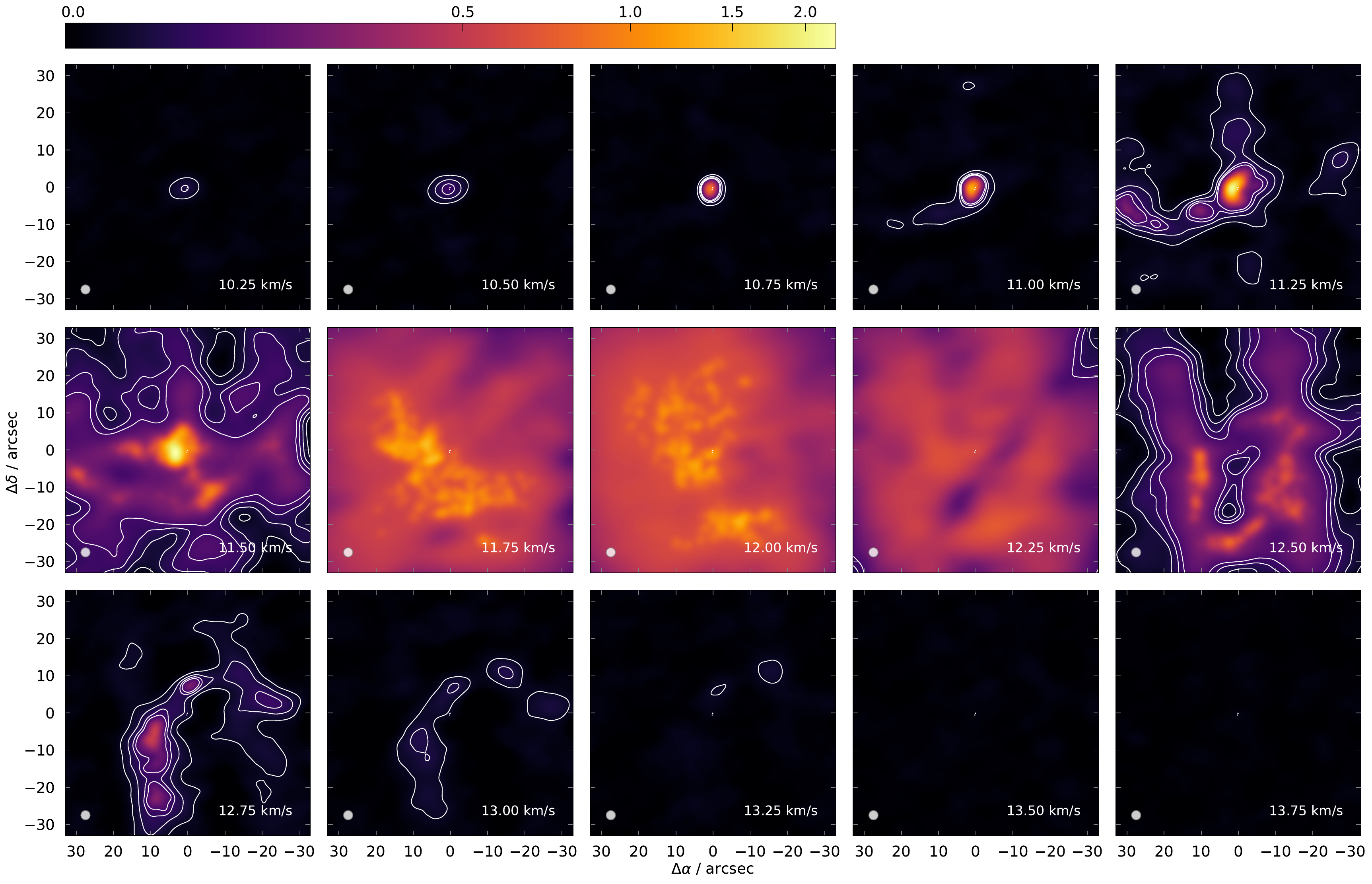}
\caption{Same as Figure~\ref{12COchan} but for $^{13}$CO.\label{13COchan}}
\end{figure*}

\begin{figure*}
\includegraphics[width=\textwidth]{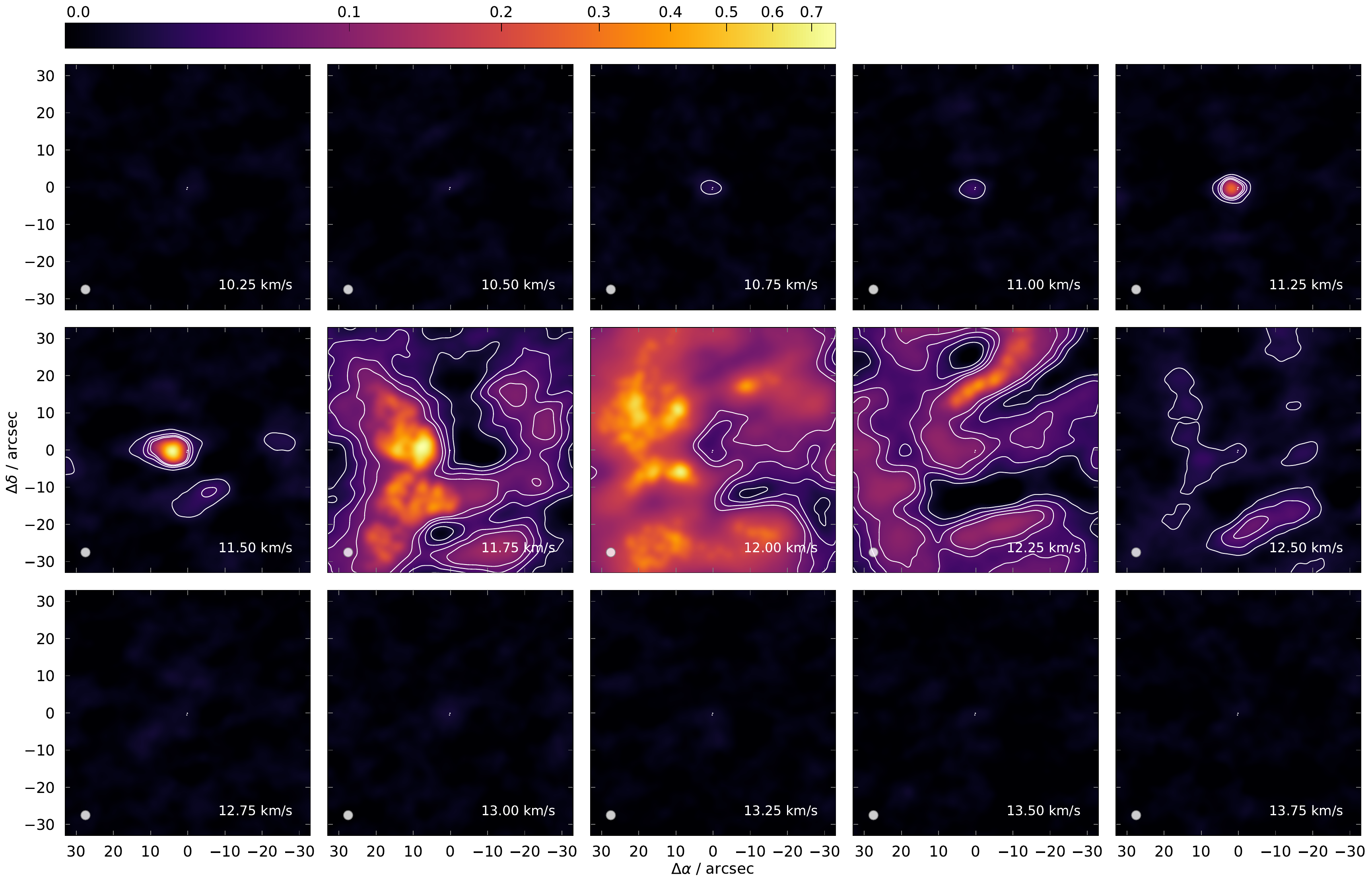}
\caption{Same as Figure~\ref{12COchan} but for C$^{18}$O.\label{C18Ochan} }
\end{figure*}

Figures~\ref{12COchan}, ~\ref{13COchan}, and ~\ref{C18Ochan} show the 12m+7m+TP array
channel maps for $^{12}$CO(2--1), $^{13}$CO(2--1), and
C$^{18}$O(2--1), respectively.


\begin{figure*}
\begin{center}
\includegraphics[angle=0,width=0.9\textwidth]{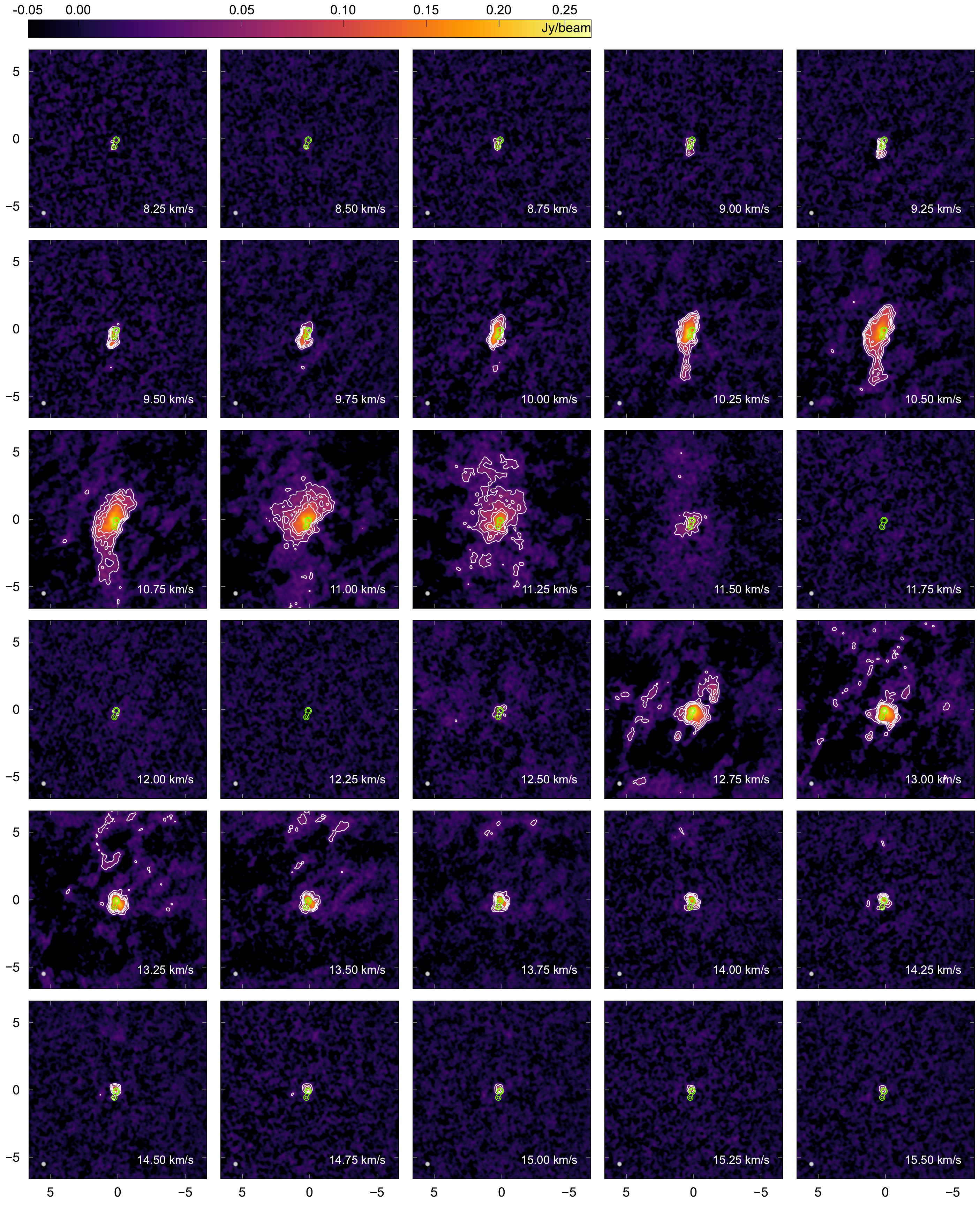}
\caption{$^{12}$CO FU~Ori maps for array C43-5 configuration.  Contours of the 1.3mm continuum emission at 0.3''
  resolution are overlaid in green (contour levels are 2 and 6
  mJy\,beam$^{-1}$).}
\label{12co_fuor_c435}

\end{center}
\end{figure*}

\begin{figure*}
\begin{center}
\includegraphics[angle=0,width=0.9\textwidth]{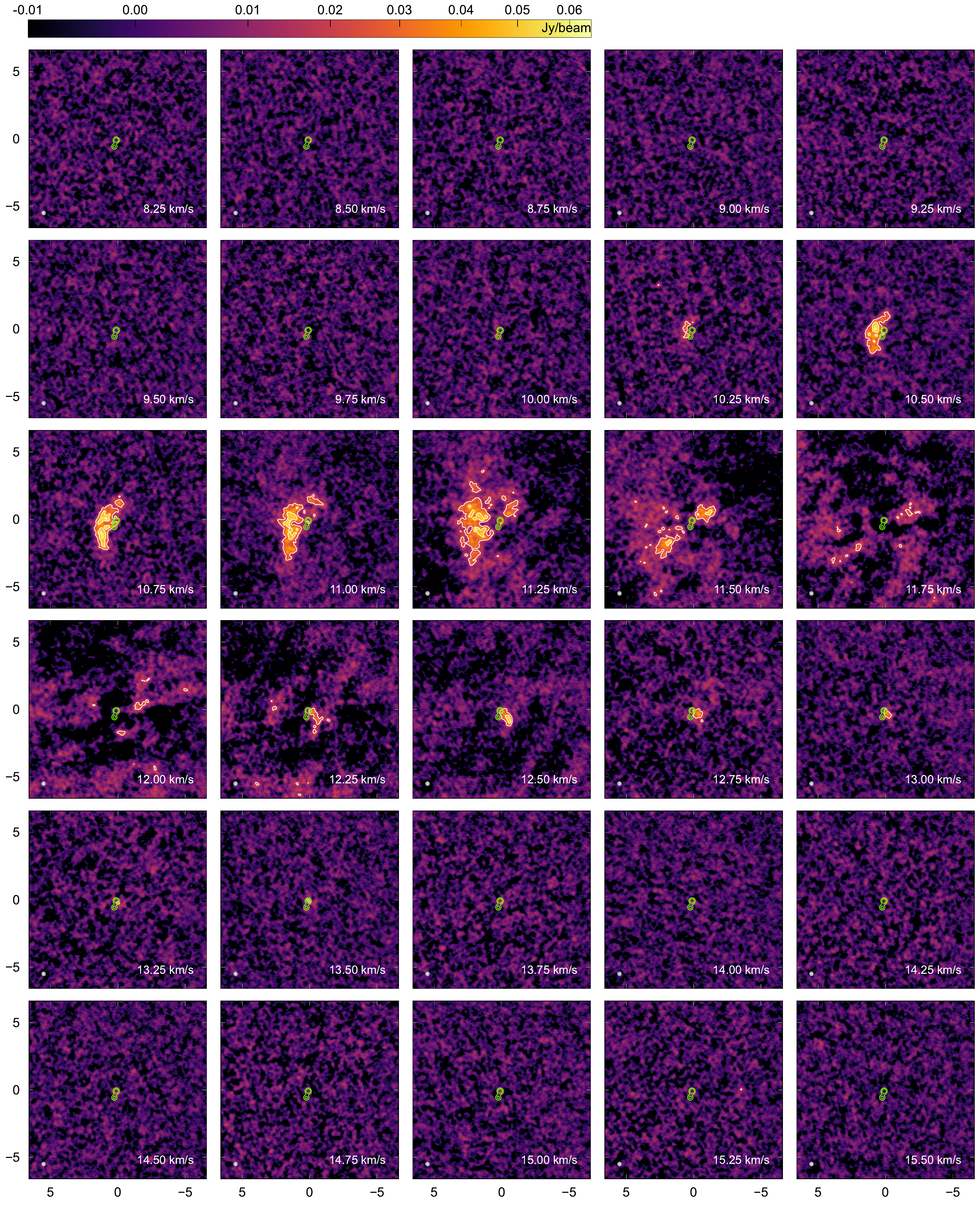}
\caption{$^{13}$CO FU~Ori maps for array C43-5 configuration.  Contours of the 1.3mm continuum emission at 0.3''
  resolution are overlaid (contour levels are 2 and 6
  mJy\,beam$^{-1}$).}
\label{13co_fuor_c435}

\end{center}
\end{figure*}

\end{document}